\documentclass[reqno,11pt]{amsart}

\usepackage{amssymb}

\begin{document}

\title[Spontaneous Symmetry Breaking and Nonlocal Regularization]
{Nonlocal regularization of abelian models with spontaneous symmetry breaking}%
\author{M. A. Clayton}%
\address{Physics Department, Acadia University, Wolfville, Nova Scotia B$0$P $1$X$0$ Canada.}%
\email{michael.clayton@acadiau.ca}%


\date{\today}

\begin{abstract}
We demonstrate how nonlocal regularization is applied to gauge invariant models with spontaneous symmetry breaking.
Motivated by the ability to find a nonlocal BRST invariance that leads to the decoupling of longitudinal gauge bosons from physical amplitudes, we show that the original formulation of the method leads to a nontrivial relationship between the nonlocal form factors that can appear in the model.
\end{abstract}
\maketitle

\section{Introduction}
The formalism for generating gauge-invariant, nonlocally regulated
actions~\cite{Evens+:1991} is a gauge-invariant application of the nonlocal
regularization pioneered by Efimov~\cite{Efimov:1967,Efimov:1968}.
It has been extended to non-Abelian models~\cite{Kleppe+Woodard:1992}, investigated beyond one-loop order and shown to treat overlapping divergences consistently~\cite{Kleppe+Woodard:1993}, and has proven to be a useful tool
when quantizing field-antifield
models~\cite{Paris:1995,Abreu:1997,Abreu:2000}. 
It has also been
applied to the investigation of anomalies in supersymmetric
models~\cite{Paris+Troost:1996,Brandt+Paris:1997}, and to the standard model in order to derive a limit on the nonlocal scale from measurements of $g-2$ for the muon~\cite{Saini+Joglekar:1997}.
This quantization procedure can be thought of as a nonlocally `deformed' (or non-canonical) quantization, and it therefore seems likely that there is a relationship between this action-level regularization and non-commutative field theories~\cite{Connes+Douglas+Schwarz:1998,Moffat:2000}.

More recently in the literature there has appeared some concern as
to how the method of nonlocal regularization is to be applied to
gauge theories, specifically to models where spontaneous symmetry
breaking is present~\cite{Basu+Joglekar:2000,Basu+Joglekar:2001}.
The issue is essentially how to fix the gauge (that is, an $R_\xi$
gauge) of a vector boson in a way that does not depend on the
gauge parameter $\xi$.

We will remain cautious in assessing the claims of gauge-dependence made in~\cite{Basu+Joglekar:2000,Basu+Joglekar:2001}.
While it superficially appears that they have not been sufficiently careful when simplifying the BRST invariance of the nonlocal theory (and therefore they do not truly have a nonlocal action that is invariant under the BRST transformation that they use when quantizing the theory), the model that they choose to work with is also not the simplest choice in which to investigate such matters.
Nevertheless, we feel that it is worthwhile to make available some results on the nonlocal regularization of gauge theories that we discovered some time ago.

In Section~\ref{sect:UB} we consider an abelian Higgs model in the
unbroken phase, showing that there is a natural choice for
smearing functions that allows a nonlocal BRST invariance with
Ward-Takahashi identities that are ``essentially'' identical to
those of the local theory. 
We then show in Section~\ref{sect:B}
that the same considerations lead to a fairly simple
nonlocalization in the broken phase of the model with the same properties.
The essence of these results lies in the treatment of the longitudinal part of the vector field propagator and its relationship with the ghost propagator.

\section{Nonlocal regularization of the unbroken model}
\label{sect:UB}

We will consider here an Abelian vector field coupled to a complex
scalar Higgs field with Lagrangian (see, for
example~\cite{Ryder:1985,Itzykson+Zuber:1980}):
\begin{equation}
L=-\frac{1}{4e^2}F^{\mu\nu}F_{\mu\nu}
 +(\partial_\mu\Phi^*+iA_\mu\Phi^*)
 (\partial^\mu\Phi-iA^\mu\Phi)
 -V(\lvert \Phi\rvert^2),
\end{equation}
which is invariant under the infinitesimal $\mathrm{U}(1)$ transformation $\delta A_\mu=\partial_\mu\theta$ and $\delta\Phi=i\theta\Phi$.
We use the standard $R_\xi$ gauge-fixing term: $(\partial_\mu A^\mu)^2/(2\xi)$, and including BRST ghost and anti-ghost fields $C$ and $\bar{C}$, which in the absence of symmetry breaking, contributes a kinetic term to the Lagrangian: $-\partial^\mu\bar{C}\partial_\mu C$.
The resulting local, gauge-fixed Lagrangian has kinetic terms
\begin{subequations}
\begin{equation}\label{eq:ub lkinetic}
K=\frac{1}{2e^2}A^\mu D^{-1}_{\mu\nu} A^\nu
 +\Phi^*D^{-1}\Phi
 -\bar{C}D^{-1}C,
\end{equation}
and higher-order interactions
\begin{equation}\label{eq:ub lint}
I=iA^\mu(\Phi^*\partial_\mu\Phi-\Phi\partial_\mu\Phi^*)
 +A^\mu A_\mu\lvert \Phi\rvert^2
-V(\lvert \Phi\rvert^2),
\end{equation}
\end{subequations}
where we have defined the inverse propagators
\begin{gather}
D^{-1}=-\square,\quad D^{-1}_{\mu\nu}=\eta_{\mu\nu}\square
-\bigl(1-\xi^{-1}\bigr)\partial_\mu\partial_\nu.
\end{gather}
The action is invariant under the BRST transformation:
\begin{equation}\label{eq:ub lBRST}
\delta A_\mu=-\partial_\mu C \zeta,\quad
\delta\Phi=-iC\Phi\zeta,\quad
\delta \bar{C}=\frac{1}{\xi e^2} \partial_\mu A^\mu\zeta,\quad
\delta C=0.
\end{equation}

\subsection{Nonlocal Regularized Lagrangian}
\label{sect:ub nonlocal action}

Nonlocal regularization of a local gauge field theory involves two
ingredients: a)~nonlocal form factors that act as a high frequency
cutoff in the loop integrals without introducing any new degrees
of freedom into the model, and b)~higher-order, nonlocal
interaction terms that allow an extended, nonlocal version of any
gauge symmetry present in the local theory.
When these two ingredients operate together we have a finite gauge field theory, with the regulator implemented directly in the Lagrangian rather than during the perturbative (loop) expansion.  We will begin here by focusing
primarily on the algebraic structure of the derived nonlocal
action that allows a nonlocally deformed gauge symmetry to
survive.

Following~\cite{Kleppe+Woodard:1992}, for every physical field $\Psi$ (which we will use throughout to symbolically represent the local fields, in this case: $A_\mu$, $\Phi$, $C$ or $\bar{C}$) we introduce a shadow field $\psi$ (in this case representing one of $a_\mu$, $\phi$, $c$ or $\bar{c}$, as above), and the `hatted' and `barred' propagators related to $\Psi$ and $\psi$ respectively.
For the scalar and ghost fields they are:
\begin{subequations}\label{eq:hat bar}
\begin{align}
\hat{D}^{-1}_\Phi&=\mathcal{E}_\Phi^{-2}D^{-1},&
\bar{D}^{-1}_\Phi&=(1-\mathcal{E}_\Phi^2)^{-1}D^{-1},\\
\hat{D}^{-1}_C&=\mathcal{E}_C^{-2}D^{-1},&
\bar{D}_C&=(1-\mathcal{E}_C^2)^{-1}D^{-1},
\end{align}
\end{subequations}
where in order that the smearing functions $\mathcal{E}$ do not introduce any additional degrees of freedom over that of the local theory, they are taken to be entire functions of the momentum operator~\cite{Efimov:1967,Efimov:1968}.
In addition, we normalize the hatted propagators to have unit residue on shell, so that since all fields are massless: $\mathcal{E}(p^2=0)=1$.  This condition guarantees that the barred propagators do not contain a pole, and therefore the shadow fields are not independent quantum degrees of freedom.
For the vector field we introduce different smearing factors for the longitudinal and transverse propagator as:
\begin{subequations}\label{eq:ADs}
\begin{align}
\hat{D}^{-1}_{\mu\nu}&=\mathcal{E}_T^{-2}
(\square\eta_{\mu\nu}-\partial_\mu\partial_\nu)
+\xi^{-1}\mathcal{E}_L^{-2}\partial_\mu\partial_\nu
,\\
\bar{D}_{\mu\nu}&=(1-\mathcal{E}_T^2)^{-1}
(\square\eta_{\mu\nu}-\partial_\mu\partial_\nu)
+\xi^{-1}(1-\mathcal{E}_L^2)^{-1}\partial_\mu\partial_\nu.
\end{align}
\end{subequations}

The quadratic terms of the nonlocal Lagrangian are introduced as
\begin{equation}
\begin{split}
K_{\mathrm{NL}}& =\frac{1}{2e^2}A^\mu \hat{D}^{-1}_{\mu\nu} A^\nu
 +\Phi^*\hat{D}_\Phi^{-1}\Phi
 -\bar{C}\hat{D}_C^{-1}C\\
 &+\frac{1}{2e^2}a^\mu \bar{D}^{-1}_{\mu\nu} a^\nu
 +\phi^*\bar{D}_\Phi^{-1}\phi
 -\bar{c}\bar{D}_C^{-1}c,
\end{split}
\end{equation}
with interaction terms identical to~\eqref{eq:ub lint} with the replacement $\Psi\rightarrow \Psi + \psi$:
\begin{equation}
I_{\mathrm{NL}}=I(A_\mu+a_\mu,\Phi+\phi,C+c,\bar{C}+\bar{c}).
\end{equation}
The nonlocal action $S_{\mathrm{NL}}=\int dx(K_{\mathrm{NL}}+I_{\mathrm{NL}})$ possesses a nonlocal version of the local BRST invariance~\eqref{eq:ub lBRST}:
\begin{subequations}\label{eq:ub BRST}
\begin{align}
\tilde{\delta} A_\mu&=-\mathcal{E}_L^2\partial_\mu (C+c) \zeta, &
 \tilde{\delta} a_\mu&=-(1-\mathcal{E}_L^2)\partial_\mu (C+c) \zeta,\\
\tilde{\delta}\Phi&=-i\mathcal{E}_\Phi^2(C+c)(\Phi+\phi)\zeta,&
\tilde{\delta}\phi&=-i(1-\mathcal{E}_\Phi^2)(C+c)(\Phi+\phi)\zeta,\\
 \tilde{\delta}\bar{C}&=\frac{1}{\xi e^2}\mathcal{E}_C^2\partial_\mu(A_\mu+a_\mu)\zeta, &
 \tilde{\delta} \bar{c}&=\frac{1}{\xi e^2}(1-\mathcal{E}_C^2)
 \partial_\mu(A_\mu+a_\mu)\zeta,\\
 \tilde{\delta} C&=0,& \tilde{\delta} c&=0.
\end{align}
\end{subequations}
By construction, these transformations satisfy $\tilde{\delta} \Psi+\tilde{\delta} \psi = \delta[\Psi+\psi]$, that is, the transformation of the field plus its shadow results in the local BRST transformation~\eqref{eq:ub lBRST} with the replacement $\Psi\rightarrow\Psi+\psi$.
This results in the interaction terms transforming in exactly the same way as in the local theory with this same replacement.
Furthermore, in the variation of the kinetic terms the smearing operators are removed, and the variation of the kinetic
terms of the field and its shadow combine to give a ``local''
result, for example:
\begin{equation}
 A^\mu \hat{D}^{-1}_{\mu\nu} \tilde{\delta} A^\nu
 +a^\mu \bar{D}^{-1}_{\mu\nu} \tilde{\delta} a^\nu
 =-(A_\mu+a_\mu) D^{-1}_{\mu\nu}\partial^\nu
 (C+c)\zeta.
\end{equation}
We see that by construction, the nonlocal Lagrangian must possess
the BRST invariance~\eqref{eq:ub BRST} as a consequence of the
BRST invariance of the local theory.

\subsection{A Modified Nonlocal BRST Invariance}
\label{sect:ub BRST}

As it stands we have a regulated nonlocal action which possesses
the BRST invariance~\eqref{eq:ub BRST}, but with a great deal of
latitude on how to choose the different functions $\mathcal{E}$.
Different choices will result in distinct nonlocal theories, all
of which will be equivalent in the local limit (it is to be
expected that higher-order vertices are generated in these
nonlocal models, but since they should vanish in the local limit
the theory would be renormalizable in that limit).

In particular, we note (following~\cite{Kleppe+Woodard:1992}) that
the BRST invariance~\eqref{eq:ub BRST} does not guarantee
decoupling of the longitudinal vector boson from $n$-point
functions. A related BRST invariance $\tilde{\delta}$ that does
(and we show this in Section~\ref{eq:ub WT}) is found by requiring
that the linear parts of the BRST transformation can be made
identical to those of the local theory~\eqref{eq:ub lBRST}. That
this is possible puts nontrivial constraints on the form of the
smearing operators--constraints that in hindsight are perhaps not
so surprising.

We begin by choosing the transformation of the anti-ghost fields
as
\begin{equation}
\tilde{\delta}_1 \bar{C} =
 \frac{1}{\xi e^2}\partial_\mu A_\mu\zeta, \quad
 \tilde{\delta}_1 \bar{c}=\frac{1}{\xi e^2}\partial_\mu a_\mu\zeta,
\end{equation}
and since this means that $\tilde{\delta}_1
\bar{C}+\tilde{\delta}_1 \bar{c}=\tilde{\delta}
\bar{C}+\tilde{\delta}\bar{c}$, the interaction terms will be
transformed in the same way by $\tilde{\delta}_1$ and
$\tilde{\delta}$. The kinetic terms for the ghosts transform into
\begin{equation}\label{eq:delta barC}
 -\tilde{\delta}\bar{C}\hat{D}^{-1}_CC
 -\tilde{\delta}\bar{c}\bar{D}^{-1}_Cc
 =\frac{1}{\xi e^2}\partial_\mu A_\mu\zeta \hat{D}^{-1}_C C
 + \frac{1}{\xi e^2}\partial_\mu a_\mu\zeta \bar{D}^{-1}_C c,
\end{equation}
and in order for this to be canceled by transforming the vector
field kinetic terms, we find that the vector field and its shadow
must transform as
\begin{equation}
\tilde{\delta}_1 A_\mu
=-\frac{\mathcal{E}_L^2}{\mathcal{E}_C^2}\partial_\mu C \zeta,
\quad
 \tilde{\delta}_1 a_\mu =-\frac{1-\mathcal{E}_L^2}{1-\mathcal{E}_C^2}\partial_\mu c
 \zeta.
\end{equation}
From this we see that only if we choose
$\mathcal{E}_C=\mathcal{E}_L$ will $\tilde{\delta}_1
A_\mu+\tilde{\delta}_1 a_\mu =\tilde{\delta} A_\mu+\tilde{\delta}
a_\mu$, and therefore the interaction term again transform the
same way under $\tilde{\delta}_1$ as they did under
$\tilde{\delta}$.

Provided we make this choice, the nonlocal action is invariant
under the modified nonlocal BRST transformation:
\begin{subequations}\label{eq:ub BRST mod}
\begin{align}
\label{eq:delta1 A}
 \tilde{\delta}_1 A_\mu&=-\partial_\mu C \zeta, &
 \tilde{\delta}_1 a_\mu&=-\partial_\mu c \zeta,\\
 \tilde{\delta}_1\Phi&=-i\mathcal{E}_\Phi^2(C+c)(\Phi+\phi)\zeta,&
\tilde{\delta}_1\phi&=-i(1-\mathcal{E}_\Phi^2)(C+c)(\Phi+\phi)\zeta,\\
 \tilde{\delta}_1\bar{C}&=\frac{1}{\xi e^2}\partial_\mu A^\mu\zeta, &
 \tilde{\delta}_1 \bar{c}&=\frac{1}{\xi e^2}\partial_\mu a^\mu\zeta,\\
 \tilde{\delta}_1 C&=0,& \tilde{\delta}_1 c&=0,
\end{align}
\end{subequations}
using which, as we shall see, decoupling is easily proven. Note
that the nonlinear parts of the transformations $\tilde{\delta}_1$
and $\tilde{\delta}$ are identical.

This is the first ingredient: the ability to make this
transformation requires that the smearing for the ghosts and the
longitudinal part of the vector field are identical. This is a
simple matter to arrange, but there is further good reason to
choose $\mathcal{E}_L$ so that it is related to $\mathcal{E}_T$.

Decomposing the vector field propagator into transverse and
longitudinal pieces, we have
\begin{subequations}\label{eq:LT}
\begin{equation}
D^{-1}_{\mu\nu}=\square T_{\mu\nu} +\xi^{-1}\square L_{\mu\nu},
\end{equation}
where the transverse and longitudinal projection operators are
\begin{equation}
T_{\mu\nu}=\eta_{\mu\nu}-\square^{-1}\partial_\mu\partial_\nu,\quad
L_{\mu\nu}=\square^{-1}\partial_\mu\partial_\nu.
\end{equation}
\end{subequations}
If we construct the smeared nonlocal kinetic terms from an entire
function of the local kinetic terms as outlined
in~\cite{Kleppe+Woodard:1992}, then we can write it as a power
series in the local kinetic terms (the constant and first-order
coefficients are constrained by the condition that there be a pole
at $p^2=0$ with unit residue):
\begin{subequations}\label{eq:vector hat}
\begin{equation}
\hat{D}^{-1}_{\mu\nu}
=\hat{D}^{-1}_{\mu\alpha}\eta^{\alpha\beta}\Bigl(\eta_{\beta\nu}
+\sum_{n=1}^\infty A_n (D^{-1})^n_{\beta\nu}\Bigr),
\end{equation}
where
\begin{equation}
(D^{-1})^n_{\alpha_0 \alpha_n}=
\prod_{m=0}^{n-1}D^{-1}_{\alpha_m\alpha_{m+1}}.
\end{equation}
\end{subequations}
With the longitudinal-transverse
decomposition~\eqref{eq:LT} this may be worked out explicitly:
\begin{equation}
(D^{-1})^n_{\mu\nu}=\square^nT_{\mu\nu}+\frac{1}{\xi^n}\square^n L_{\mu\nu},
\end{equation}
so that
\begin{equation}
\hat{D}^{-1}_{\mu\nu}=\Bigl(1+\sum_{n=1}^\infty A_n\square^n\Bigr)\square T_{\mu\nu}
+\Bigl(1+\sum_{n=1}^\infty \frac{A_n}{\xi^n}\square^n\Bigr)
\frac{1}{\xi}\square L_{\mu\nu},
\end{equation}
and by comparison with~\eqref{eq:ADs} we see that it is natural to choose
\begin{equation}\label{eq:LT relation}
\mathcal{E}_L(\square)=\mathcal{E}_T\Bigl(\frac{1}{\xi}\square\Bigr).
\end{equation}

Accepting this condition, we see that we only have the freedom to
choose two smearing operators: $\mathcal{E}_T$ and
$\mathcal{E}_\Phi$--the smearing function for the ghost is
constrained by requiring the existence of the BRST
symmetry~\eqref{eq:ub BRST mod}, and the smearing functions for
the longitudinal and transverse related by~\eqref{eq:LT relation}.
In fact, if we follow the method of nonlocal regularization as formulated in~\cite{Kleppe+Woodard:1992} to the letter, we should really be considering entire functions of the operator appearing in the quadratic part of the local action as a whole.  That is, writing~\eqref{eq:ub lkinetic} as $K=\Psi D^{-1}\Psi$, then we should write $\hat{D}^{-1}=\sum_n A_n D^{-1}$.
This leads to $\mathcal{E}_\Phi=\mathcal{E}_T$, and we have a nonlocalization that depends on a single nonlocal form factor function only.

\subsection{Quantization and Ward-Takahashi Identities}
\label{eq:ub WT}

Quantization of the nonlocal theory proceeds in two steps.
First, because the propagators for the shadow fields do not possess a pole in the propagator, the shadow field equations:
\begin{equation}\label{eq:shadow feq}
\psi=-\bar{D}\frac{\delta I_{\mathrm{NL}}}{\delta \psi},
\end{equation}
constitute in implicitly-defined, but local relationship between the shadow fields and physical fields $\Psi$.
The quantum nonlocal BRST action is generated by iteratively replacing the shadow fields in $S_{\mathrm{NL}}$ using~\eqref{eq:shadow feq}, which generates the nonlocal Lagrangian as a series which is straightforward to evaluate to any order.
Similarly, the resulting action will be invariant under the nonlocal BRST transformations of the local fields $\Psi$ as given by $\tilde{\delta}$ or $\tilde{\delta}_1$, with the same replacement of the shadow fields using~\eqref{eq:shadow feq}.

Once this procedure is completed, path integral quantization is completed through the prescription to compute the vacuum expectation of any operator $\mathcal{O}$ via (see~\cite{Evens+:1991} for a discussion of the $\mathcal{T}^*$ ordered product)
\begin{equation}
\langle \mathcal{T}^*[\mathcal{O}]\rangle
 =\int d\mu_{\mathrm{inv}}\mathcal{O}\exp(iS_{\mathrm{NL}}).
\end{equation}
The measure $d\mu_{\mathrm{inv}}$ is an invariant measure that can
be written as
\begin{equation}
d\mu_{\mathrm{inv}}=d\Psi
\exp(iS_{\mathrm{meas}}),
\end{equation}
where $S_{\mathrm{meas}}$ may be determined from the BRST transformations (up to BRST-invariant contributions) via the condition 
\begin{equation}\label{eq:meas determination}
\delta S_{\mathrm{meas}}=i\mathrm{Tr}
 \Bigl[\frac{\partial}{\partial \Psi}\tilde{\delta}_1 \Psi\Bigr];
\end{equation}
see~\cite{Evens+:1991,Kleppe+Woodard:1992} for a more complete discussion on
this matter.

We have explicitly used the modified nonlocal BRST transformation $\tilde{\delta}_1$ in~\eqref{eq:meas determination}, since generating the measure from this transformation will lead to a nonlocal quantum theory will be invariant under this transformation.
Note though, that beginning from different nonlocal gauge symmetries we
would generate a different invariant measure factor, and therefore
a different nonlocal quantum theory and different Ward-Takahashi identities.
Also, since $\mathcal{E}_L$ is $\xi$-dependent, we have to expect that
the measure will also have nontrivial dependence on the gauge
parameter.

To generate the Ward-Takahashi identities for the nonlocal quantum theory, we introduce the generating functional
\begin{equation}\label{eq:GF}
\mathcal{Z}_J=\int d\mu_{\mathrm{inv}}\exp(iS_{\mathrm{NL}}+iJ\cdot\Psi),
\end{equation}
where
\begin{equation}
 J\cdot\Psi=\int dx\;[J^\mu A_\mu + J\Phi + \bar{C}D +\bar{D}C
  +iU\mathcal{E}_\Phi^2(C+c)(\Phi+\phi)],
\end{equation}
and we have added a complex source $U$ for the BRST invariant
field combination: $i\mathcal{E}_\Phi^2(C+c)(\Phi+\phi)$ (to see this, note
that $(C+c)^2=0$). Following~\cite{Ryder:1985}, in order for the generating functional to result in a BRST invariant
perturbation theory, it must itself be gauge invariant, and
transforming $\mathcal{Z}_J$ using~\eqref{eq:ub BRST mod} we end up with the
condition
\begin{equation}
\int dx\Bigl[
 J^\mu\partial_\mu \frac{\delta \mathcal{Z}_J}{\delta \bar{D}}
 +\frac{1}{\xi e^2}\partial_\mu\frac{\delta \mathcal{Z}_J}{\delta J^\mu}D
 +J\frac{\delta \mathcal{Z}_J}{\delta U}
 \Bigr]=0.
\end{equation}
This leads to the Ward-Takahashi identities on the vertex
functional $\Gamma[\psi]$, which is related to $\mathcal{Z}_J$ by
$\mathcal{Z}_J=\exp(i\Gamma+iJ\cdot\psi)$:
\begin{equation}
\int dx\Bigl[
 \frac{\delta \Gamma}{\delta A^\mu} \partial_\mu C
 +\frac{1}{\xi e^2}\frac{\delta \Gamma}{\delta \bar{C}}\partial_\mu A^\mu
 +i\frac{\delta \Gamma}{\delta \Phi}\mathcal{E}_\Phi^2(C+c)(\Phi+\phi)
 \Bigr]=0.
\end{equation}
These conditions are identical to those of the local theory except
for the nonlocal vertices appearing in the final term. In
particular, taking functional derivatives with respect to $C(y)$
and $A_\nu(z)$ leads to the relationship between the ghost
two-point function $\Pi_C$ and the longitudinal projection of the
vector field two-point function: $\xi e^2\partial_\mu
\Pi^{\mu\nu}+\partial^\nu \Pi_C=0$, which, since no ghosts can
exist on external legs, shows that the physical vector field
propagator is transverse. Further functional derivatives with
respect to the vector field guarantee that the longitudinal vector
field decouples from all physical pure vector field $n$-point
functions.

The first manifestation of nonlocality in the Ward-Takahashi
identities is in the relationship between the scalar field vacuum
expectation value $\langle\Phi\rangle$ and the longitudinal part
of the two-point mixing of the vector field and the scalar field
$\Pi^\mu$: $\partial_\mu\Pi^\mu
=i\mathcal{E}^2_\Phi\langle\Phi\rangle$. In the local theory this
relation indicates how a real vacuum expectation value for the
scalar field will result in a mixing between the Goldstone boson and the longitudinal part of
the vector field, resulting in a mass for the vector field. That
this relation involves the scalar field smearing function
$\mathcal{E}^2_\Phi$, foreshadows that in the spontaneously broken theory we will be forced to choose the smearing
function for the scalar field to be related to that of the vector
field if we want the modified BRST invariance to exist for the nonlocal theory.

\section{Broken symmetry phase}
\label{sect:B}

It is noteworthy that spontaneous symmetry breaking
necessarily mixes the quadratic and interacting terms--the masses
are `fed' down from the interaction terms to the quadratic terms
by the shift to the potential minimum. 
Since the nonlocal regularization treats the kinetic and interaction terms
differently, it is not a straightforward matter to go from the
nonlocally-regularized model in the previous section to a
regularized model in the broken phase. 
Instead we will begin with the BRST invariant local theory in the broken phase, and apply the nonlocal regularization method as outlined in Section~\ref{sect:UB}.

Assuming a form of the scalar field potential that leads to spontaneous symmetry breaking:
\begin{equation}
 V(\lvert \phi\rvert^2)
 =\lambda\bigl(\lvert\phi\rvert^2-\tfrac{1}{2}v^2\bigr)^2,
\end{equation}
we redefine the scalar field fluctuations about the minimum of this
potential as
\begin{equation}
\phi=2^{-\frac{1}{2}}(v+H+iF),
\end{equation}
and at the same time introduce the gauge-fixing and ghost
contributions:
\begin{equation}
L_{\mathrm{gf}}=
-\frac{1}{2\xi e^2}(\partial_\mu A^\mu+\xi e^2 v F)^2,\quad
L_{\mathrm{ghost}}=
 -\partial^\mu\bar{C}\partial_\mu C +\xi e^2v (v+H)\bar{C}C.
\end{equation}
The resulting BRST Lagrangian has quadratic terms
\begin{subequations}
\begin{equation}\label{eq:local quadratic}
K=\frac{1}{2e^2}A^\mu D^{-1}_{A,\mu\nu}A^\nu
+\frac{1}{2}HD^{-1}_HH +\frac{1}{2} F D^{-1}_{F} F
 -\bar{C}D^{-1}_{C}C,
\end{equation}
and higher-order interactions
\begin{equation}\label{eq:local int}
\begin{split}
I &= A^\mu( F\partial_\mu H-H\partial_\mu F)
 +vH(A^\mu A_\mu -\lambda F^2-\lambda H^2) \\
 &+\tfrac{1}{2}A^\mu A_\mu(H^2+ F^2)
 -\tfrac{1}{4}\lambda(H^2+ F^2)^2
 +\xi e^2 v H\bar{C}C.
\end{split}
\end{equation}
\end{subequations}
In~\eqref{eq:local quadratic} we have integrated by parts and
defined the inverse propagators:
\begin{gather}
D^{-1}_H=-(\square+2\lambda v^2),\quad
D^{-1}_F=D^{-1}_C=-(\square+\xi e^2 v^2),\\
D^{-1}_{A,\mu\nu}=\eta_{\mu\nu}(\square+e^2v^2)
-\bigl(1-\xi^{-1}\bigr)\partial_\mu\partial_\nu,
\end{gather}
and action is invariant under the local BRST transformations
\begin{subequations}\label{eq:BRST}
\begin{gather}
\delta A_\mu=-\partial_\mu C\zeta, \quad \delta
F=-C(v+H)\zeta,\quad
\delta H=C F \zeta,\\
\delta C=0,\quad \delta\bar{C} = \frac{1}{\xi e^2}(\partial_\mu
A^\mu+\xi e^2v F)\zeta.
\end{gather}
\end{subequations}

\subsection{Nonlocal Regularized Lagrangian}
\label{sect:regularization}

The construction goes through as described in Section~\ref{sect:ub
nonlocal action}: we introduce shadow fields for all local fields,
make the $\Psi\rightarrow\Psi+\psi$ replacement in the interaction
terms, and introduce the hatted and barred propagators for the
scalar and ghost fields of the same form as~\eqref{eq:hat bar}
\begin{equation}
\hat{D}^{-1}_{\Psi}=\mathcal{E}_\Psi^{-2}D^{-1}_{\Psi},\quad
\bar{D}^{-1}_{\Psi}=(1-\mathcal{E}_\Psi^2)^{-1}D^{-1}_{\Psi}.
\end{equation}
Once again we allow the possibility that the smearing functions
are different for the different fields, except that now the
propagator for the Higgs field $H$ will now have a pole at
$p^2=2\lambda v^2$, and so it is natural to choose
$\mathcal{E}_H=\mathcal{E}_H(\square+2\lambda v^2)$.

Noting that the local propagator for the vector field can be written as
\begin{equation}
 D^{-1}_{A,\mu\nu}
 =(\square+e^2v^2)T_{\mu\nu}
 +\xi^{-1}(\square+\xi e^2v^2)L_{\mu\nu},
\end{equation}
then the same argument that led to~\eqref{eq:LT relation} leads us to choose
\begin{equation}\label{eq:LTB relation}
\mathcal{E}_T=\mathcal{E}_T(\square+e^2v^2),\quad
\mathcal{E}_L=\mathcal{E}_T\bigl(\xi^{-1}\square+e^2v^2\bigr),
\end{equation}
and the propagators for the vector field and its shadow are:
\begin{subequations}
\begin{align}
\hat{D}^{-1}_{A,\mu\nu}
&=\mathcal{E}_T^{-2}(\square+e^2v^2)T_{\mu\nu}
+\mathcal{E}_L^{-2}\xi^{-1}(\square+\xi e^2v^2)L_{\mu\nu},\\
 \bar{D}^{-1}_{A,\mu\nu}&=(1-\mathcal{E}_T^2)^{-1}(\square+e^2v^2)T_{\mu\nu}
 +(1-\mathcal{E}_L^2)^{-1}
 \xi^{-1}(\square+\xi e^2v^2)L_{\mu\nu}.
\end{align}
\end{subequations}
Note that the vector field propagator has a physical pole at $p^2=e^2v^2$ as well as the gauge-dependent pole at $p^2=\xi e^2v^2$, both of which are reflected in the smearing functions.

The quadratic terms in the nonlocal Lagrangian are:
\begin{equation}
\begin{split}
K_{\mathrm{NL}}&= \frac{1}{2e^2}A^\mu \hat{D}^{-1}_{A,\mu\nu}A^\nu
+\frac{1}{2}H\hat{D}^{-1}_HH +\frac{1}{2} F \hat{D}^{-1}_{F} F
+\bar{C} \hat{D}^{-1}_{C}C\\
&+\frac{1}{2e^2}a^\mu \bar{D}^{-1}_{A,\mu\nu}a^\nu
+\frac{1}{2}h\bar{D}^{-1}_Hh
+\frac{1}{2} f \bar{D}^{-1}_{F} f
+\bar{c} \bar{D}^{-1}_{C}c,
\end{split}
\end{equation}
with higher-order interaction terms determined as before from the local interaction terms~\eqref{eq:local int}: $I_{\mathrm{NL}}=I(\Psi+\psi)$, and the nonlocal BRST action is invariant under the nonlocal version of the local BRST symmetry~\eqref{eq:BRST}:
\begin{subequations}\label{eq:broken BRST}
\begin{gather}
\tilde{\delta} A_\mu=-\mathcal{E}_L^2\partial_\mu (C+c)\zeta,\\
\tilde{\delta} C=0,\quad
\tilde{\delta}\bar{C} = \mathcal{E}_C^2\frac{1}{\xi e^2}
(\partial_\mu (A^\mu+a^\mu)+\xi e^2v (F+f))\zeta,\\
\tilde{\delta}  F=-\mathcal{E}_F^2(C+c)(v+H+h)\zeta,\quad
\tilde{\delta} H=\mathcal{E}_H^2(C+c)(F+f) \zeta,
\end{gather}
\end{subequations}
with shadow field transformations that follow the pattern given in~\eqref{eq:ub BRST}.

\subsection{A Modified Nonlocal BRST Invariance}

As described in Section~\ref{sect:ub BRST}, we want to find a
modified nonlocal BRST transformation $\tilde{\delta}_1$ in which
the linear part of the transformation is identical to the linear
part of the local BRST transformation~\eqref{eq:BRST}, and the
nonlinear part identical to the nonlinear part of~\eqref{eq:broken
BRST}. Requiring that the ghosts transform as:
$\tilde{\delta}_1\bar{C}=(\frac{1}{\xi e^2}\partial_\mu A^\mu + v
F)\zeta$, then transforming the quadratic terms for the vector
field we find that we have to choose
$\mathcal{E}_C=\mathcal{E}_L$, and the vector field would
transform as~\eqref{eq:delta1 A}.  Also transforming the quadratic
terms of the Goldstone boson we find that we have to choose
$\mathcal{E}_F=\mathcal{E}_C$, and the linear part of the BRST transformation of the
Goldstone boson $F$ is also `localized'. The result of this is the
modified nonlocal BRST invariance:
\begin{subequations}
\begin{gather}
\tilde{\delta}_1  F=-vC\zeta-\mathcal{E}_L^2(C+c)(H+h)\zeta, \\
\tilde{\delta}_1 H =\mathcal{E}_H^2(C+c)(F+f) \zeta, \quad
\tilde{\delta}_1 A_\mu =-\partial_\mu C\zeta, \\
\tilde{\delta}_1\bar{C} = \Bigl(\frac{1}{\xi e^2}\partial_\mu
A^\mu
 + v F\Bigr)\zeta, \quad
\tilde{\delta}_1 C=0,
\end{gather}
\end{subequations}
with the ghosts transforming following the pattern in~\eqref{eq:ub
BRST mod}.

Note that choosing $\mathcal{E}_L=\mathcal{E}_C=\mathcal{E}_F$ does not interfere with the shadow
fields $f$, $c$ and $\bar{c}$ being removable at the classical
level, since their `barred' propagators still do not contain a pole.
This would not necessarily been the case with an
arbitrary $\mathcal{E}_L$, but would be possible with any $\mathcal{E}_L=\mathcal{E}_L(\square+\xi e^2 v^2)$ that satisfies $\mathcal{E}_L(0)=1$.
Nevertheless, the relation~\eqref{eq:LTB relation} is well-motivated, and leaves only the freedom to choose $\mathcal{E}_T$ and $\mathcal{E}_H$.
If we further require the theory to generated from a single entire function of the kinetic terms of the local theory, as described at the end of Section~\ref{sect:ub BRST}, the we would also be forced to choose $\mathcal{E}_H(x)=\mathcal{E}_T(x)$, that is, $\mathcal{E}_H$ is the same function of $\square+2\lambda v^2$ that $\mathcal{E}_T$ is of $\square+e^2v^2$.

\subsection{Quantization and Ward-Takahashi Identities}

Path integral quantization proceeds as before, except that in the generating functional~\eqref{eq:GF} we will write the source terms as
\begin{multline}
 J\cdot\Psi=\int dx\;[J^\mu A_\mu + JF + KH + \bar{C}D +\bar{D}C
 \\
  +U\mathcal{E}_H^2(C+c)(F+f)
  -V\mathcal{E}_L^2(C+c)(H+h)],
\end{multline}
and proceeding as before, we find that the vertex functions will satisfy:
\begin{multline}
\int dx\Bigl[
 \frac{\delta \Gamma}{\delta A^\mu}\partial_\mu C
 +vC\frac{\delta \Gamma}{\delta F}
 +\Bigl(\frac{1}{\xi e^2}\partial_\mu A^\mu +vF\Bigr)\frac{\delta \Gamma}{\delta
 \bar{C}}\\
 +\frac{\delta \Gamma}{\delta F}\mathcal{E}_H^2(C+c)(F+f)
 -\frac{\delta \Gamma}{\delta H}\mathcal{E}_L^2(C+c)(H+h)
 \Bigr]=0.
\end{multline}
As before, these lead to the decoupling of the longitudinal vector boson from $n$-point functions that involve external vector fields.

\section{Discussion}
\label{sect:discussion}

We have presented a simple prescription for nonlocal
regularization of field theory models with spontaneous symmetry
breaking in an arbitrary $R_\xi$ gauge.
The smearing functions depend on the gauge parameter in a nontrivial way, and we therefore expect that the path integral measure factor will also depend on $\xi$.
Nevertheless, requiring that the nonlocal theory possess a nonlocal BRST invariance with Ward-Takahashi identities that imply the decoupling of longitudinal gauge bosons, leads to a nontrivial relationship between the nonlocal form factors for different fields.
This relationship was shown to follow naturally from the formalism presented in~\cite{Kleppe+Woodard:1992}, which results in a nonlocal theory that depends on a single nonlocal form factor.

Although in hindsight these relationships are not surprising, recent related constructions have appeared in the literature~\cite{Basu+Joglekar:2000,Basu+Joglekar:2001} do not impose them.
While one cannot say that alternate constructions are incorrect, one can make the case that the method presented herein is preferred on the grounds of simplicity of implementation and interpretation.


\end{document}